\documentclass[11pt]{article}
\usepackage[margin=1.0in]{geometry}
\usepackage{graphicx}
\usepackage{epsfig}
\usepackage{amssymb,amsfonts,amsmath,amsthm}
\usepackage{mathrsfs}
\usepackage{epstopdf}
\usepackage{hyperref,color}
\usepackage{natbib}
\usepackage{setspace}
\usepackage[multiple]{footmisc}

\pdfoutput=1

\begin{document}

\title{Quantum gravity from general relativity}
\author{Christian W\"uthrich\thanks{I owe thanks to Karen Crowther, Nick Huggett, Niels Linnemann, and Tushar Menon for comments on earlier drafts. This work was partly performed under a collaborative agreement between the University of Illinois at Chicago and the University of Geneva and made possible by grant number 56314 from the John Templeton Foundation and its content are solely the responsibility of the author and do not represent the official views of the John Templeton Foundation.}}
\date{28 November 2017}
\maketitle

\noindent
{\small Forthcoming in Eleanor Knox and Alastair Wilson (eds.), {\em Routledge Companion to the Philosophy of Physics}.}

\begin{abstract}\noindent
Although general relativity is a predictively successful theory, it treats matter as classical rather than as quantum. For this reason, it will have to be replaced by a more fundamental quantum theory of gravity. Attempts to formulate a quantum theory of gravity suggest that such a theory may have radical consequences for the nature, and indeed the fate, of spacetime. The present article articulates what this problem of spacetime is and traces it through three approaches to quantum gravity taking general relativity as their vantage point: semi-classical gravity, causal set theory, and loop quantum gravity.
\end{abstract}

\section{Introduction: the need to go beyond general relativity}\label{sec:intro}

General relativity (GR) is our best theory of gravity and one of the best confirmed theories in the history of science \citep{wil14}. Why then do physicists believe that it needs to be replaced by a more fundamental theory of `quantum gravity'?\footnote{Nota bene that this is not equivalent to {\em quantizing gravity}, since a theory that can deal with both quantum effects of matter and relativistic phenomena, i.e., a theory of quantum gravity, may be achieved by other means.} Standard arguments brought forth for such a need may be motivational, but can hardly be considered conclusive. 

First, there is unification. But that the world conforms to this metaphysical ideal is certainly not an a priori truth, although there is some inductive evidence that unification is methodologically valuable in physics. Furthermore, a common interpretation of GR has it that GR shows that gravity is not a force, and hence not similar in kind to the three fundamental forces---weak and strong nuclear, and electromagnetic.\footnote{A less common interpretation of GR views gravity as the result of a massless spin-2 field and thus much more like a force.} Given the distinction in kind, any need for unification becomes much less pressing. A second argument derives from Eppley and Hannah's thought experiment \citep{epphan77} and tries to establish not only that a quantum theory of gravity is needed, but that, additionally, it must be obtained by quantizing gravity. However, the argument contains loopholes---and the thought experiment seems physically impossible \citep{hugcal01,wut05,mat06}. Another argument, explicated e.g.\ by \citet{dopeal95} but originating earlier, claims that since no localization of a physical system by means of a detection by a photon can be made with an accuracy exceeding the Planck length, as this is the scale at which a photon, if compressed to a spacetime volume smaller than that scale, would collapse into a black hole and thus become unusable for detection services. However, this argument, even if successful, at best establishes that spacetime is operationally discrete in that there are limits beyond which we cannot detect its structure. It does manifestly not imply anything about the structure of spacetime itself \citep{wut05}. As a fourth example, \citet{perter01} have argued that a theory cannot accommodate classical and quantum interactions, on pain of a violation of central physical principles---specifically the correspondence principle. Their argument relies on an articulation of the problem in Koopmanian terms, which combine classical and quantum physics, and thus remains hostage to a particular formalism that is far from compulsory \citep{wut05}. 

Instead of proceeding from a presumption that these quantum considerations invade and in some sense invalidate GR, one could argue by identifying a fatal flaw in GR itself. The most popular target for this strategy is the fact that GR predicts the existence of singularities, which flag a breakdown of the mathematical representation as used in the theory. In light of the singularity theorems, which establish that under rather generic conditions, cosmological models of GR are singular, it looks as if we cannot dismiss singular spacetimes as mere unphysical artifacts; they are here to stay with GR. While it is reasonable to operate methodologically on the assumption that nature affords a regular mathematical description, there is of course no metaphysical guarantee that limitations to such a description are not objective features of the world. Perhaps the world just isn't amenable to scientific study in this way. 

Similarly, one might construct an argument from an apparent insufficiency of thermodynamics: black holes potentially violate the second law, as we can conceivably drop an entropy package into a black hole such that this entropy then vanishes behind the hole's event horizon and is thus `lost' to the total entropy of the exterior universe. Consequently, we ought to generalize the second to account for the potential loss \citep{bek73}. The simplest way to do that is to attribute an entropy to black holes, such that any such potential loss is at least made up by an increase of the hole's entropy. From statistical mechanics, we have come to expect that any physical system with an entropy has `microstates'. Hence, black holes have microscopic states such that `entropy' becomes an applicable property. Thus, contrary to what classical GR asserts, black holes have additional properties beyond their mass, charge, and angular momentum. Therefore, the argument concludes, we need to go beyond GR to understand the micro-structure of black holes, i.e., their `quantum nature'. However, this argument falls short on two accounts. First, it presupposes, perhaps incorrectly, the validity of the second law in cases where black holes are involved. Second, Bekenstein's attribution of entropy to black holes inappropriately depends on a fundamentally information-theoretic approach to physically---specifically when it infers to the existence of micro-structure from the existence of entropy \citep{wut18}. Black holes may well be thermodynamic objects; but if this is so, it has to be for reasons other than those given by Bekenstein. If they are thermodynamic, they may require a quantum theory to describe the microstates that give rise to their entropy. As suggestive as this line of reasoning may be, it falls short from being conclusive.

However, even if all these arguments do not conclusively establish that GR needs to be supplanted by a more fundamental theory of quantum gravity, there is a single reason which does: it is a contingent, but extremely well established fact that matter is quantum---not classical as GR assumes. For this reason alone, GR cannot be the last word on gravity and will eventually have to be replaced by a theory of gravity that incorporates the fact that matter is quantum. None of this has, prima facie, any implications for the structure of spacetime. However, as it turns out, the most promising approaches to a formulation of such a theory all have more or less radical consequences for spacetime. 

Now with the need to go beyond GR established, there are various ways of reaching beyond GR. The next section (\S\ref{sec:semi}) briefly discusses a natural and conceptually simple extension of GR---semi-classical gravity---, which basically plugs quantum fields into the right-hand side of the Einstein field equation. \S\ref{sec:cst} then briefly introduces the first of the two approaches to full quantum gravity discussed here, causal set theory. The remainder of this essay is devoted to canonical quantum gravity and particularly to its main representative, loop quantum gravity, which will be introduced in \S\ref{sec:lqg}. \S\ref{sec:spt} is concerned with what I will call `the problem of spacetime' in the context of loop quantum gravity (see Th\'ebault, this collection, for a review of the problem of time in canonical quantum gravity). There are of course other approaches to quantum gravity, most of which do not consciously start from GR and are thus not the subject of the present essay (see e.g.\ Dawid, this collection).

\section{Semi-classical gravity}\label{sec:semi}

The most straightforward way to a theory of quantum gravity is by means of the {\em semi-classical approach to quantum gravity}. Semi-classical gravity treats gravity classically in the sense that it uses the framework of GR to describe, and assumes that the matter fields propagating in spacetime are quantum fields, described by an appropriate quantum field theory (QFT). The two are coupled to one another through the {\em semi-classical Einstein field equation}:
\begin{equation}
R_{ab} - \frac{1}{2} g_{ab} R = 8\pi \langle\psi|\hat{T}_{ab}|\psi\rangle,
\end{equation}
where $\langle\psi|\hat{T}_{ab}|\psi\rangle$ is the expectation value of the stress-energy tensor of the quantum fields in a physically reasonable state $|\psi\rangle$. Already by the time Wald's classic {\em Quantum Field Theory in Curved Spacetime and Black Hole Thermodynamics} \citep{wal94} appeared almost a quarter century ago, this approach could offer a fully satisfactory and mathematically rigorous theory for linear quantum fields in curved spacetime (p.\ 1, the construction follows in ch.\ 4).

However, semi-classical gravity faces severe difficulties and limitations. First, there is an ambiguity in the definition of $\langle\psi|\hat{T}_{ab}|\psi\rangle$, which could be resolved by a more fundamental theory, or be fixed by experiment \citep[98]{wal94}.\footnote{Cf.\ \citet{ver12} for a more recent, and more optimistic review. As it turns out, the undetermined renormalization parameters may serve to fix the vacuum energy, and thus solve the `dark energy' problem in cosmology \citep{dapeal08}.} Although this speaks not against the {\em truth} of semi-classical gravity, it is a sign of its non-fundamentality. Second, as $\langle\psi|\hat{T}_{ab}|\psi\rangle$ contains terms of fourth order in derivatives of the metric (and not just of second order), and the semi-classical Einstein equation will have new solutions, often with `runaway' character \citep[99]{wal94}. Third, standard energy conditions can be generically violated under physically reasonable conditions in semi-classical gravity \citep{cur16a}. However, these point-wise energy conditions can arguably be replaced by weaker counterparts serving a similar purpose, the so-called `quantum energy conditions', which merely prohibit that the energy density can be arbitrarily negative over long enough periods of time. As it turns out, these weaker conditions are generally not violated in semi-classical gravity (Fewster review). Finally, it is sometimes listed as a problem of the approach that it is generally impossible to compute $\langle\psi|\hat{T}_{ab}|\psi\rangle$. Although true, such more practical limitations befall all approaches to quantum gravity, and much of physics besides. For instance, the Navier-Stokes equations are of great conceptual utility, and arguably at least approximately true, even though their general solution is not known. In fact, the {\em classical} Einstein equation remains generally unsolved after a century.\footnote{I am grateful to Rainer Verch for discussions on these points.} Nevertheless, at least the first three kinds of technical difficulties may well reflect a deeper physical tension in the approach.

Recently, what has become known as the `firewall paradox' can be thought of as a new challenge to semi-classical gravity arising from black-hole physics. If the argument by \citep{almeal13} leading to the firewall paradox is accepted, then either (i) the dynamical evolution from matter falling into the black hole to outgoing Hawking radiation is not unitary, or (ii) a form of the equivalence principle is not true, or (iii) the usual semi-classical approach of QFT on (slightly) curved spacetime is not valid. The `paradox' is interesting because each of the options forces us to give up what appears to be an eminently reasonable and successful assumption behind well-confirmed physics. It is truly, as Raphael Bousso puts it, ``a menu from hell'' \citep{oue12}. While most physicists, including the authors of the original article, seem to favour discharging the equivalence principle into retirement---hence the moniker `firewall' as event horizons would then burn up infalling observers---, it may thus be the case that semi-classical gravity in the sense introduced here is not valid. 

No doubt semi-classical gravity deserves more philosophical scrutiny than it has so far received. But one may well ask---and not just in the light of the foregoing challenges---, how a semi-classical m\'elange of physical principles could possibly justify that quantum physics and gravity are blended into a unified fundamental theory when the latter is generally expected to reject at least some of the dearly held principles on which the former is built. All this may indicate, as most physicists think, that semi-classical gravity is confined to nothing but a small, temporary, and incomplete extension of `old physics', and that therefore a bolder approach to quantum gravity is required, at least as a fundamental theory. It is the purpose of the remainder of this article to introduce two such attempts to articulate a quantum theory beyond GR with ambitions to be offer a more fundamental account. Common to these approaches is that they both take GR as their vantage point for quantum gravity.

\section{Causal set theory}\label{sec:cst}

Causal set theory (CST) \citep{bomeal87} takes the central insight of GR to be that, for causally sufficiently non-pathological relativistic spacetimes, the causal structure of a spacetime determines its full geometry up to a conformal factor \citep{mal77}.\footnote{For a recent introductory survey, see \citet{dow13}.} In a popular slogan in the field, `spacetime = causality  + size'. This insight motivates taking the fundamental structures underlying relativistic spacetimes to be discrete causal sets, where a fundamental relation of causal connectibility underwrites the causal structure, and the discreteness fixes the conformal factor, i.e., it provides `size' information in the form of countable elementary, Planck-sized `chunks' of spacetime. A causal set is represented by an ordered pair of a set of elementary events and a binary relation $\prec$ of `causal precedence' which partially orders the set of elementary events. That the relation $\prec$ gives rise to a partial ordering of the basal events means that it is reflexive, antisymmetric, and transitive. Its antisymmetry rules out certain causal pathologies that famously afflict GR, such as the presence of closed timelike curves. The elementary events have no further physical properties beyond standing in relations of causal precedence with other events. Finally, the resulting structure is stipulated to be discrete. The CST research program has so far not delivered a \emph{quantum} theory of gravity, and so remains a work in progress. The following discussion is thus confined to the classical theory.

The central question for CST is whether the postulated causal sets generically give rise to relativistic spacetimes, or an empirically indiscernibly close surrogate thereof. As stated in the previous paragraph, this is demonstrably not the case: almost all realistically large causal sets as defined above form so-called `KR orders' consisting of only three `generations' of elements with about half the elements in the middle generation and a quarter each in the first and last generations, giving us no realistic model of the history of the universe. Thus, additional conditions need to be imposed in order to restrict the set of admissible causal sets such that physically realistic models come to dominate. If this were accomplished, then  causal set theory could justifiedly claim to offer a viable, though still classical, theory of the structure underlying relativistic spacetimes. In this sense, additional, `dynamical' conditions are imposed. The most widely discussed approach is the classical sequential growth dynamics as introduced by \citet{ridsor99}, which is based on the remarkable result that a small number of physically justifiable assumptions severely constrain the possible dynamics if the latter is understood as the totally ordered sequence of the `birthing' of elements accreting to a past-finite, future-infinite causal set. As there is nothing quantum about this proposed dynamical condition---it is a thoroughly classical prescription---, classical sequential growth dynamics is offered as a stepping stone to a full quantum theory of gravity.

A central part of the proposal is that the total order of the sequence of `birthing' is not a physical aspect of the model, but rather an auxiliary construction to obtain the right kinds of causal sets, i.e., those that will generically give rise to past-finite relativistic spacetimes. It thus remains an undisputed option to interpret the resulting causal sets in the usual eternalist fashion honoured by relativistic physics. However, leading causal set theorists \citet{sor07} and \citet{dow14} have argued that this dynamics can be interpreted in A-theoretic terms, i.e., involving an ineliminable and substantive notion of passage, and thus vindicates a metaphysics of becoming compatible with relativistic physics. The metaphysics favoured by Sorkin and Dowker most closely resembles that of a growing block. The price to be paid for an A-theoretic metaphysics, however, is that the relativistically kosher becoming cannot be global, i.e., spatially extended, on pain of violating the Lorentz symmetry that any relativistic theory must accommodate. Instead, we have what Sorkin dubbed an ``asynchronous multiplicity'' of localized becoming \citeyearpar[158]{sor07} as, metaphorically, in a tree which independently grows at the tips of its different branches such that it is meaningless to say that one tip objectively grew before the other. As \citet[358n]{poo13} has correctly noted, this view is close to Fine's \citeyearpar{fin05} ``non-standard A-theory'', which rejects that there are, fundamentally, absolute tensed facts, but instead insists that there are fundamental tensed facts which are relativized to inertial frames. Consequently, fundamental reality is fragmented. 

Naturally, the arguments of Sorkin and Dowker have been taken up by philosophers, who have examined the claimed compatibility of a metaphysics involving a substantive, A-theoretic notion of becoming with relativistic physics \citep{but07,ear08,wutcal17}. It turns out that the usual dilemma foisted on an A-theoretic metaphysics by relativistic physics \citep{cal00,wut13} essentially survives into causal set theory \citep{wutcal17}: any notion of an objective, global becoming either answers to the A-theorist's explanatory request, or is compatible with Lorentz symmetry, but not both. The way in which the dilemma is resolved in the Sorkin-Dowker interpretation of causal set dynamics is in accepting that the objective sense of becoming or of a present is not global, but only local---`asynchronous'. However, as it turns out, there remains the possibility of a bizarre metaphysics of a growing block with global becoming: whole swathes of the sum total of existence remain, sometimes for long periods on the cosmological clock, in an ontological indeterminate, liminal state between existence and non-existence \citep{wutcal17}. 

The main attraction of CST is that it offers, based on a central insight in GR, an ontologically clear picture---at least as long as one sticks to an eternalist, B-theoretic reading of its metaphysics and to a classical version of its dynamics. Unfortunately, this very achievement is at peril by the necessary transposition of the theory's main tenets into a quantum theory, which remains, as mentioned above, unaccomplished.

\section{Loop quantum gravity}\label{sec:lqg}
Another approach to developing a full quantum theory of gravity by starting out from GR subjects it to a so-called `canonical quantization' procedure.\footnote{For a recent and accessible introduction, see \citet{rovvid15}, which also covers the covariant extensions of the theory. REF TO RICKLES, this volume?} This is generally a promising strategy, as the procedure has been applied to classical theories and successfully delivered effective quantum theories on other occasions. However, the application of canonical quantization is less straightforward in GR, as the fact that GR treats spacetime as a four-dimensional unit cannot be easily reconciled with the presupposition of canonical quantization that physical theories deal with three-dimensional systems which dynamically evolve over time. In order to compensate, as it were, for the required split of the four-dimensional structure of relativistic spacetime, constraints on the basic Hamiltonian variables arise. The resulting constraint equations are equivalent to the Einstein field equation; since the theory must presuppose a topology permitting a global time function, it is classically only equivalent to a part of GR and does therefore not fully capture it. 

The first important choice for any programme of quantization along these lines is to select a pair of canonical variables. A first promising set of variables based on the four-metric $g_{ab}$ was proposed by \citet{adm62}: the three-metric $q_{ij}$, the `lapse' function $N = \sqrt{-g^{00}}$, and the `shift' function $N_i = g_{i0}$, where $a,b = 0,...,3$ are spacetime indices and $i, j = 1, 2, 3$ are merely spatial indices. The great advantage of these `ADM' variables is their intuitive geometric interpretation. The three-metric is the metric tensor induced by the four-metric on three-dimensional constant-$t$ spacelike hypersurfaces, the lapse function represents the proper time elapsed between the $t$-hypersurface and the $(t+dt)$-hypersurface, and the shift function measures the displacement of the spatial coordinates between these two hypersurfaces in stationary coordinates. Unfortunately, this approach leads to a dauntingly hard form of the constraint equation, and progress has stalled along these lines. 

A more promising canonical quantization re-expresses the geometry of classical relativistic spacetimes in terms of the connection $A_a^i$ and its conjugate, a densitized triad $E_i^a$, rather than the metric-based ADM variables. These basic, so-called `Ashtekar variables' are then used to construct a `holonomy-flux algebra', which consists of holonomies of the basic connection, which use parallel transport around closed loops as a measure of the curvature of the connection, and its conjugate, fluxes constructed from the densitized triads. One then seeks a representation of this algebra in some appropriate Hilbert space and expresses the constraint equations in therms of the holonomies and the fluxes, which are then defined as operators on that Hilbert space. The elements of that Hilbert space which satisfy the constraint equations would then be those which represent the physically possible quantum states of the gravitational field. The `physical Hilbert space' would be the Hilbert space consisting of all and only those states which are physically possible in this sense. 

Unfortunately, however, the so-called Hamiltonian constraint equation, thought to capture the dynamical content of the theory, has so far resisted its solution. The subspace consisting of those states, which satisfy all the other constraints, forms a Hilbert space as well and is known as the `kinematical Hilbert space $\mathcal{H}_K$. At it turns out, there exists a useful orthogonal basis of $\mathcal{H}_K$ which permits a natural physical interpretation. The elements of this basis are the so-called `spin network states', which are the eigenstates of two important geometric operators, the `area' and `volume' operators. The spin network states in the basis are built up from a state, which can be interpreted to represent the quantum geometric vacuum, by iterated application of the holonomies as `creation' operators which raise the excitation level of the fluxes. 

The spin network states are normally represented as graphs, which carry spin representations on their edges and vertices. The structure of the graph gives the network of adjacency relations between parts of the spin network and thus represents the `connectivity' of the basic structure. The spin representations are related to the eigenvalues of the geometric operators applied to parts of these spin networks. The representations sitting on the vertices are the eigenvalues of the volume operator, and those on the edges the eigenvalues of the area operator. These eigenvalues have a discrete spectrum with a non-zero, positive smallest value. Thus, the spin network basis lends itself to a natural physical interpretation in terms of geometric properties and suggests a fundamentally granular quantum structure that gives rise, in a yet to understand classical limit, to the smooth spacetimes of GR. The spin representation of the vertices then gives a measure of their volume, and the representations on the edges connecting two adjacent vertices a measure of the area of their connecting `surface'. 

The spin network states do not, in general, solve the Hamiltonian constraint equation and are thus merely `kinematical'. Consequently, they are routinely interpreted as `spatial' in that they are what underlies three-dimensional physical space as an aspect of four-dimensional spacetime. It thus appears as if at least the structure of manifest space might be straightforwardly explicable on this approach as a fundamentally discrete structure of granular parts, which combine by attaching to one another through their adjacency relations in a manner of Lego-like building blocks. Unfortunately, such view would be too simplistic and cannot be maintained. First, even at the kinematic level, we should not forget that generically, the state of the gravitational field is a {\em quantum superposition} of elements of the spin network basis of $\mathcal{H}_K$. Thus, even if the basis element afford an interpretation in straightforward terms, the generic state is one that superposes basis states of different, inconsistent geometric properties and so does not yield to a geometric interpretation. Second, in order to get the full ontological picture, the Hamiltonian constraint equation would have to be solved and the fully physical Hilbert space would have to be known. 

As the canonical quantization procedure has thus encountered a formidable roadblock, physicists are seeking ways to circumvent the problem. They pursue two general strategies. The first simplifies the systems studied by the theory and thus reduces the numbers of degrees of freedom already at the classical level and then attempts a standard canonical quantization on the simplified theory. This succeeds, and it leads to what is interpreted to give us the cosmological sector of the theory---Loop Quantum Cosmology---, as the symmetry-reduced classical system is isotropic and homogeneous.\footnote{For an introduction to Loop Quantum Cosmology, cf.\ \citet{boj11}.} As such, it should give us insight into the very early universe from a fundamental perspective, and hopefully solve at least some of the many challenges of present-day cosmology. Advocates claim that Loop Quantum Cosmology does indeed offer such insight: it is alleged to dissolve the initial singularity (`big bang') of the standard model of cosmology (\citet[Ch.\ 7]{boj11}, but see also \citet[Ch.\ 6]{wut06}), and to lead to an early inflationary period with a graceful exit without the additional postulation of an inflaton field or similar ad hoc stipulations \citep{boj02}. The general idea behind this research problem is now to successively introduce complexities such as anisotropies in a climb to a more realistic full theory. To what extent this endeavour will succeed is open.

The second avenue pursued by physicists is to abandon the canonical path half way through the procedure and replace the dynamics with a covariant approach, as advocated e.g.\ in \citet{rovvid15}. Instead of the Hamiltonian $\hat{H}$ of the canonical approach, the dynamics is expressed, hopefully equivalently, in terms of the corresponding transition amplitudes. In a quantum theory, the transition amplitude from an initial state to a final state in the form of a probability capture the dynamical content of a theory, as is computed as an integral over paths. In the absence of a spacetime background for a quantum system to propagate along `paths', the connection between an `initial' and a `final' spin network state is made through combinatorial `splittings', `persistings', and `joinings' of the granular structure, adorned with a probability for each. The resulting fundamental structure, which is what is assumed to ground classical spacetime, is a so-called `spinfoam'. 

Although many questions remain open, such as how to incorporate general matter fields into the picture, spinfoams have promising advantages. First, and in an echo to Loop Quantum Cosmology, one can also formulate a theory of quantum cosmology on this basis \citep[\S11.3]{rovvid15}. Second, the spinfoam approach permits a `derivation' of the Bekenstein-Hawking formula for the entropy of black holes \citep[\S10.4]{rovvid15}. 

\section{The problem of spacetime}\label{sec:spt}

Putting open problems and past accomplishments of LQG to the side, how should one conceive of the resulting fundamental structure, to the extent to which it is currently known and understood? In particular, how non-spatiotemporal is it, and how can classical smooth spacetime re-emerge from it? This section will address these philosophically central issues.\footnote{For a recent review, cf.\ \citet{mat17}.}

As LQG is based on a quantization procedure which presupposes a foliation of spacetime into three-dimensional spaces totally ordered by a one-dimensional time, the destinies of space and time are not entirely parallel. Let us start with time. Like all other canonical approaches to quantum gravity, LQG suffers from a `problem of time'.\footnote{See also the contribution by Th\'ebault to this volume and \citet[\S2]{hugeal13}, and references therein.} This problem has two aspects. First, since the Hamiltonian operator $\hat{H}$, which generates the dynamics, also turns out to be a constraint (in the approach based on ADM variables as well as in LQG), we obtain as the basic dynamic equation for a physical state $|\Psi\rangle$
\begin{equation*}
\hat{H} |\Psi\rangle = 0.
\end{equation*}
Since only the states $|\Psi\rangle$ which satisfy this constraint equation can be considered physically possible, it seems as if a physical state cannot change over time. All its truly physical properties, it seems, must be represented by operators which commute with the Hamiltonian and are thus constants of motion. Genuinely physical properties cannot change over time. This first aspect of the problem of time thus really is a {\em problem of change}. The second aspect of the problem of time at the quantum level is that in all approaches for which we have an explicit expression for $\hat{H}$, there is no time: it appears as if time, as a physical quantity, has simply fallen by the wayside. Although the problem of time requires much more careful scrutiny than it can be given here, it leaves us with the puzzling issue of how an apparently fully temporal world, buzzing and beaming with change, can emerge from what appears to be a fundamentally fully `static' or `frozen' structure. 

Space also undergoes a change from GR to LQG, though nowhere near as complete as time seems to. As we noted in \S\ref{sec:lqg}, the fundamental spin network states are certainly discrete and thus lack some of the structure of relativistic spacetime. However, there is more: the fact that smooth physical space is supposed to arise from a quantum superposition of the geometric spin network states such that the resulting state has no determinate geometric properties surely stands in need of explanation. Clearly, the quantum measurement problem rears its hydra head here once more, but the problem seems even deeper now that we are not dealing with a particle having no determinate position even though we always detect it somewhere, but instead with space(-time) itself not having any determinate geometric properties although we never fail to experience it in any other way. So how can it be that we have determinate and measurable geometric information of spacetime at our scales when the fundamental structure will not generally have corresponding geometric properties? 

Based on \citet{butish99} and as explained in more detail in \citet{wut17a}, an answer to this difficult question consists of two parts. First, one needs to identify the quantum states with approximately `classical', i.e.\ geometric, properties and articulate a physical mechanism that `drives' the system toward those states. These semi-classical states are thought to correspond to almost flat three-spaces with at most small quantum fluctuations. A promising way of identifying them among the states in $\mathcal{H}_K$ is the `weave state' approach using coherent states \citep{ashrovsmo92}. These weave states are (almost) eigenstates of the `volume' operator. This operator earns its name by virtue of the fact that its eigenvalues approximate the corresponding classical values for the three-volume of a region in spacetime as determined by the classical gravitational field. Moreover, these same states are (almost) eigenstates of the `area' operator, which corresponds to the classical property of the area of the two-surface of a spacetime region. The selection of weave states stands in need of justification, which is far from automatic \citep[\S4.2]{wut17a}. Such justification would be delivered, for instance, by a physical mechanism which systematically drives the kinematic states to the semi-classical weave states. Decoherence, with an appropriate partition of the system's degrees of freedom into `salient'  and `background' degrees of freedom. 

The second step then consists in relating the weave states to the classical spacetimes. This will involve a limiting procedure, establishing the precise sense in which the quantum states approximate the relevant geometric properties of the emergent spacetime at sufficiently large scales. None of this yet solves the quantum measurement problem; but at least in gives us a template for how to understand the relation between the fundamental non-spatiotemporal structure and relativistic spacetimes. 

This way of relating the fundamental with the emergent is thought to be broadly reductive, and hence the notion of `emergence' at play cannot be the non-reductive concept typically used in philosophy. Rather, it designates a relation expressing the novelty of the emergent vis-\`a-vis the fundamental that is nevertheless ontologically grounded in the latter in a way that is consistent with reduction. Emergence as defined by novelty and robustness, i.e., by non-fundamental behaviour which is robust under irrelevant changes on the fundamental level, and which is logically independent of reduction, as articulated by \citet{but11a,but11b} and as developed by \citet{cro15,cro16} captures the relevant sense of emergence at stake. 

In CST, there also exists a sketch of how one can recover aspects of spacetimes from the fundamental causal sets.\footnote{For the details of this, the reader is advised to consult \citet[Ch.\ 3]{hugwut} and \citet[\S3]{lamwut18}.} Here, just as in the case of LQG, it is not all aspects of relativistic spacetimes which are obtained in such recovery: for instance, the continuum is not recovered in either case, it is merely approximated, and qualitative aspects of `spatiality' and `temporality' perhaps remain lost. However, just as statistical mechanics does not precisely recover all of thermodynamics, it is not necessary to regain all aspects of classical spacetimes; it suffices to show how the fundamental structure, be it spin network or causal sets or whatever else, can play the relevant functional roles of spacetime, such as that of spatiotemporal localization or other empirically determinable geometric properties such as distances and durations. \cite{lamwut18} develop this functionalist strategy towards an understanding of the emergence of spacetime in the cases of CST and LQG and argue that the recommended functionalist attitude rejects the necessity to regain anything beyond the empirically salient functions of spacetime. In particular, any insistence on some allegedly irreducible spacetime `qualia' is considered deeply misguided; and attempts to constitute spacetime from elementary spatiotemporal building blocks of a primitive ontology is thought to be altogether unnecessary. In this sense, some qualitative features of spacetime may well be emergent in the stronger sense, i.e., in that they cannot be reduced to what is described by a quantum theory of gravity. But this is not a loss. The one and only assignment that must be completed by any candidate theory of fundamental physics is to show how the aspects of spacetime necessary to support the physics of our manifest world arise from the structures postulated by the theory. Make no mistake: this is a formidable task, which no programme in quantum gravity can claim to have discharged to date.

\bibliographystyle{plainnat}
\bibliography{/Users/christian/Professional/Bibliographies/quantumgravity}

\end{document}